\documentclass{article}

\usepackage{authblk}
\usepackage{natbib}
\usepackage{multirow}
\usepackage{tabularx}
\usepackage{makecell}
\usepackage{hyperref}
\usepackage[margin=1in]{geometry} 
\usepackage{amsmath,amsthm,amssymb}
\usepackage{wasysym}
\usepackage[scr]{rsfso}
\usepackage{titling}
\usepackage{rotating}

\setcitestyle{authoryear,open={[},close={]}}

\usepackage{graphicx} 
\graphicspath{{figures/}} 
\usepackage[font=small,labelfont=bf]{caption}
\usepackage{wrapfig}

\usepackage{color}

\definecolor{dkred}{rgb}{0.7,0,0.1}

\providecommand{\keywords}[1]
{\small	\textbf{\textit{Keywords---}} #1}

\renewenvironment{abstract}
 {\small
  \begin{center}
  \bfseries \abstractname\vspace{-.5em}\vspace{0pt}
  \end{center}
  \list{}{
    \setlength{\leftmargin}{2cm}%
    \setlength{\rightmargin}{\leftmargin}%
  }%
  \item\relax}
 {\endlist}
 
\begin{document}

\section*{}
\title{{Logjams are not jammed: \\
measurements of log motions in Big Creek, Idaho}}

\author{Nakul S. Deshpande  \thanks{now at: Department of Earth and Environmental Sciences, University of Pennsylvania 251 Hayden Hall, 240 S 33rd St, Philadelphia, PA 19104 nakuld@sas.upenn.edu}}

\author{Benjamin T. Crosby}

\affil{\footnotesize Department of Geosciences, Idaho State University, 921 S 8th Ave, Pocatello, ID 83209}

\renewcommand\Affilfont{\itshape\small}

{\let\newpage\relax\maketitle}

\begin{abstract}
Colloquially, a ``logjam'' indicates a kinematic arrest of movement. Taken literally, it refers to a type of dense accumulation of wood in rivers widely recognized as bestowing numerous biological and physical benefits to the system but also present serious hazards to infrastructure. Despite this, no \textit{in-situ} field measurements have assessed the degree of arrest in a naturally-formed logjam. Using time-lapse photography, repeat total station surveys and water level loggers, we provide an unprecedented perspective on the evolution of a logjam in central Idaho. Despite the namesake, we find that the logjam is not jammed. The ensemble of logs progressively deforms in response to shear and buoyant lift of flowing water, modulated by the rising limb, peak and falling limb of the snowmelt hydrograph. As water rises and log drag against the bed and banks decreases, they collectively translate downstream, generating a heterogeneous pattern of deformation. As streamflow recedes and the logs reconnect with the bed and banks, the coherent deformation pattern degrades as logs settle opportunistically amongst their neighbors. Field observations of continuous movement at a low rate are qualitatively similar to creep and clogging, behaviors that are common to a wide class of disordered materials. These similarities open the possibility to inform future studies of environmental clogging, wood-laden flows, logjams, hazard mitigation and the design of engineered logjams by bridging these practices with frontier research efforts in soft matter physics and granular rheology.
\end{abstract}\hspace{10pt}

\keywords{logjams, lag rafts, jamming, clogging, anisotropic granular materials}

\pagebreak

\section{Wood in rivers}
Wood is considered both friend and foe in river systems. Accumulations of wood are critical agents in facilitating river function and form by means of rich interactions between wood, sediment, flow and ecology \citep{Gurnell2002,Faustini2003,Montgomery2003,Gurnell2005}, which render them a popular tool in river restoration efforts \citep{Gerhard2000,Kail2007,Abbe2011,Roni2014}. These efforts are in light of, and a reaction to, legacy logging practices and river corridor management which have altered wood-river feedbacks \citep{Wohl2014a}. Historically, these large accumulations were the subject of great public interest due to their prodigious size and the removal efforts that accompanied them \citep{Humphreys1971,McPhee1987} and were central in timber industry's log drives \citep{Rajala2010}. In a contemporary context, logs collected by floods can accumulate and persist at bridge piers and increase the chances of structural damage, flooding and scour, presenting hazards to communities and infrastructure \citep{Diehl1997,Lyn2007,Comiti2012,Mazzorana2009,Wohl2016}. In both cases, these accumulations are often assigned the generic descriptor `logjams' - a reference to both the material and their (apparently) physically arrested state. Many taxonomies of wood accumulations exist and depend on a number of environmental variables \citep{Wohl2010,Collins2012} and transport regime \citep{Kramer2017}. Logjam studies have been conducted in steep, single-threaded mountainous settings \citep{Wohl2014c} and low-gradient alluvial settings where wood-floodplain interactions are important \citep{Wohl2013, Collins2012}. Logjams which grow to a size such that they span the entire channel width and are of significant length are referred to as 'log rafts' -- these features can occur in low-gradient alluvial settings \citep{Boivin2015}, behind reservoir dams \citep{Moulin2004,LeLay2007,Fremier2010} and along coastlines \citep{Kramer2015}. To date, no studies have explored large, channel-spanning `log rafts' in steep, mountainous, single-threaded reaches.

Determining the full risks and beneficial potential of logjams and log rafts is contingent on an understanding of their physics, complimented by observations in the field. Current models of wood transport mechanics determine hydrodynamic entertainment thresholds for single log pieces \citep{Braudrick2000,Alonso2013}. Log-log (ensemble) movement has been observed in congested modes of transport \citep{Braudrick1997}, when multiple wood pieces are mobilized, but are little explored. In both cases, the type of transport (single log or ensemble) is strongly coupled to the channel geometry and supply mechanism in the field; in the single-log scenario, wood is supplied via near-bank recruitment and deposition \citep{Piegay2015}, while delivery via breached logjams or mass wasting and hillslope processes \citep{Wohl2009,RuizVillanueva2014} generally lead to congested modes of transport. 

Observations of wood and transport are difficult to collect in the field and few studies exist. Workers have conducted multi-year field surveys \citep{Wohl2008}, collected photos with ground cameras \citep{Benacchio2017}, tagged logs with RFID tags and GPS \citep{Ravazzolo2014,Schenk2014,Macvicar2012}. These field observations encode signatures of hysteresis \citep{Piegay2015}, are dependent on diverse fluvial process such as ice break-out floods \citep{Boivin2015} and are highly sensitive to hydrograph shape and timing \citep{RuizVillanueva2014} - all of which place strong non-linear constraints on physical frameworks. Challenges in our knowledge of congested modes of transport, coupled with the opportunity provided by growing repositories of image data and recognition of non-linearity in wood transport have motivated recent efforts in casting dense, ensemble log transport as $`$wood-laden-flows' \citep{Ruiz-Villanueva2019}. Inspired by descriptions of fluid-particle geophysical flows \citep{Iverson2005}, this framework casts ensemble kinematics of logs as a function of the corresponding wood volume-fraction of the flow, in kinship to the phenomenology of yield stress fluids. 

\section{Clogging and jamming}
Although the namesake \textit{logjam} is common in the vernacular, \textit{jamming} theory accounts for how numerous disordered, `soft' materials \citep{deGennes1992} transition from fluid to solid-like states \citep{Liu2010,Ohern2003}. These transitions also occur when particles are driven through confined geometries and clog \citep{Zuriguel2014,Thomas2015} or disordered landscapes, where obstacles inhibit movement by pinning particles that would otherwise be mobile \citep{Peter2018,Reichhardt2017}. Granular materials fall within the jamming paradigm \citep{Majmudar2007} and clogging behavior is observed in dense flows of elongated grains \citep{Borzsonyi2016,Torok2017}. Logjams have been explicitly citepd as macroscopic manifestations of granular phenomenology \citep{Borzsonyi2013} and invoked as an inspiration in the assembly of nanomaterials \citep{Whang2003,Yang2003}. Further, fruitful advances in our understanding of landscapes have been made via an embrace of the granular roots of sediment transport \citep{Zimmermann2010,Frey2011,Houssais2015,Ferdowsi2017a,Ferdowsi2018a}. Likewise, we look to these branches of physics as an inspiration for the framing and interpretation of the work herein and for informing future investigations of the physics of logjams.


\section{Field Setting and Logjam Formation}
\begin{figure}[ht]
\includegraphics[width=1\linewidth]{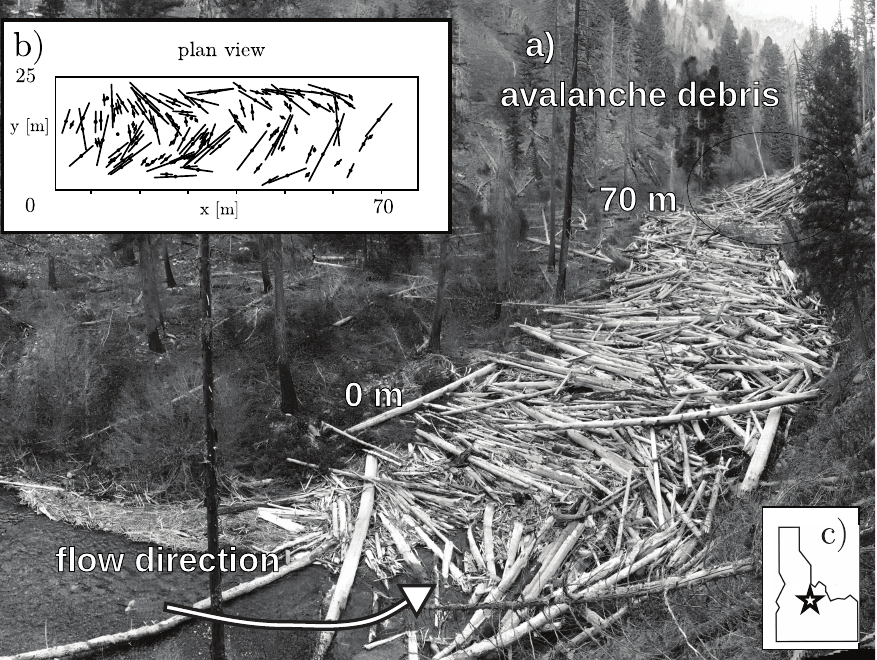}
\caption{\label{fig:location} \textbf{Big Creek logjam.} a) Photograph of studied logjam, looking downstream. Note the spatial extent of logs, the herringbone structure of the logjam, and hillslope avalanche debris (shaded region). b) Coordinate system and rendering of a sub-sample of measured log positions (lines) and centroids (dots). c) Location of Big Creek in Idaho, USA.}
\end{figure}

Big Creek is a steep, mountainous tributary of the Middle Fork Salmon River in central Idaho that drains 1540 km$^2$ and spans 1030 to 2900 m in elevation. The watershed has experienced multiple episodes of intense wildfire \citep{Arkle2010}. On March 10, 2014, a rain-on-snow event initiated a series of snow avalanches in the upper reaches of Big Creek (Supplemental Materials), delivering thousands of burned, standing dead Douglas Fir and Lodgepole Pine into the channel. By the late spring, multiple mixed ice-debris jams had blown out and floated logs downstream, their progression halted by a persisting avalanche debris deposit.  Over the next year, upstream wood progressively coalesced into a roughly thousand-member, channel spanning logjam, 70 meters long and 20 meters wide with a herringbone pattern of log orientations (Figure \ref{fig:location}). At the point of this logjam, Big Creek drains 275 km$^2$. Anecdotal reports from other researchers, USFS employees and backcountry pilots indicate that $\sim$ 20 such channel-spanning logjams presently exist within the Salmon River watershed. 

\section{Measuring Stage and Logjam Kinematics}
Following initial reconnaissance and method development in the summer and fall of 2015, the positions of $\sim$132 of indexed log tips were measured on seven occasions during the snowmelt hydrograph of May and June, 2016 (Figure \ref{hydrograph} ,Supplemental Materials Table 1). Survey dates were selected heuristically and not over fixed time intervals. Each total station survey ($\sim$ 5mm precision) was referenced to a network of stable local control points, enabling precise survey-to-survey comparisons. 3D coordinates of log tips were used to find log midpoints, from which we calculated the magnitude and direction of displacement vectors in both the xy (horizontal) and z (vertical) direction, for each log between each survey interval. In addition to the xyz coordinates of indexed log tips, we measured water surface elevations throughout the logjam at the beginning of each survey. To track stream stage during the field campaign, we installed a barometrically-corrected water-pressure transducer on a bedrock bank near the center of the logjam, recording stage ever 15 minutes. Time-lapse videos from three Moultrie M-1100i game cameras installed in trees at the upstream, central and downstream ends of the logjam enabled qualitative evaluation of deformation at 30 minute intervals, supplementing the survey and stage data.

\section{Results}

\subsection{Water level}
Two long-wavelength snowmelt peaks superimposed with lesser diel melt fluctuations characterize the flow conditions during the survey campaign (Figure 2a). River stage almost peaks in mid-May but the highest and most persistent flows occur in early June. Water surface profiles measured within the logjam slope gently and linearly downstream (Figures 3c,3d). The shape of these profiles do not change during the survey period, although they shift in elevation in concert with the hydrograph. At $\sim$50 m along the logjam, near the debris blockage, there is a pronounced inflection where the water surface drops about 0.5 m. The existence, location and magnitude of the water surface inflection persists throughout the field campaign.

\begin{figure*}[h]
\includegraphics[width=1\linewidth]{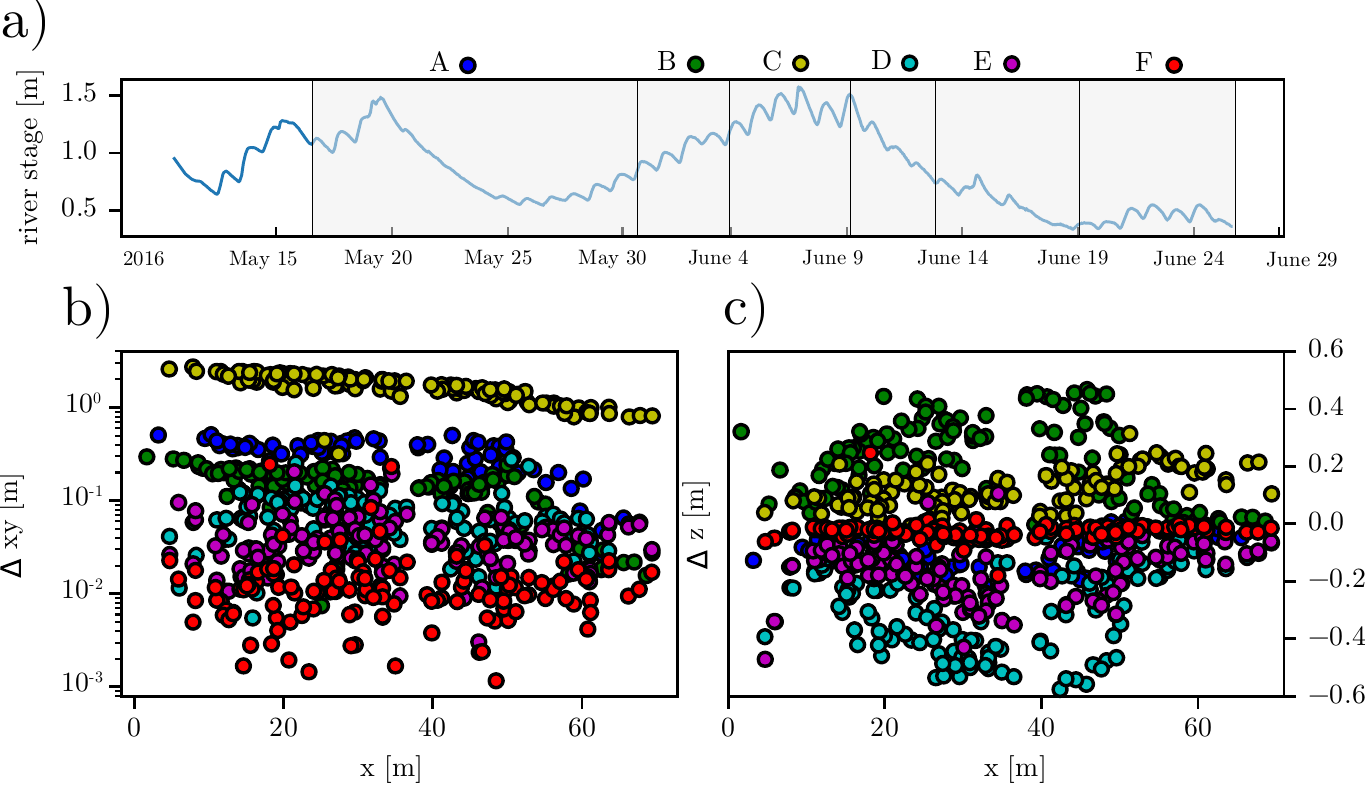}
\caption{\label{hydrograph} \textbf{Variation in river stage drives logjam deformation} a) Measurements of river stage (blue line) during the survey campaign during May-June of 2016. Shaded domains indicate the duration of measurement surveys and are labeled with the corresponding label and color. Longitudinal profile of log centroid displacements in the (b) horizonal and (c) vertical directions. Centroids are colored by survey interval. Note that horizontal centroid displacements are greater in the upstream section of the logjam and decrease downstream.}
\end{figure*}

\subsection{Total Station}
Total station data capture temporally and spatially heterogeneous patterns of deformation. Displacement distributions of are broad and tens of centimeters in magnitude during intervals A and B (Figure 2 3). The largest displacements occurred over interval C, when displacements average an order of magnitude greater than the two previous intervals. During the falling limb of the hydrograph (intervals D, E and F), displacements are low in magnitude but finite and within our measurement error. On June 9th (the beginning of interval C), log elevations cluster tightly at the water surface. During interval C, deformation proceeds uniformly in the downstream direction. Horizontal movement is greatest upstream (0-30 m) and is $\sim$3 times larger than displacements downstream (Figure 3a,3e). This is in contrast to log positions on June 19th (the beginning of interval F, during low flow), which are broadly distributed above the water surface (Figures 3b,3f). During interval E, log displacements are low in magnitude and semi-random in orientation.

\begin{figure}[h]
\includegraphics[width=1\linewidth]{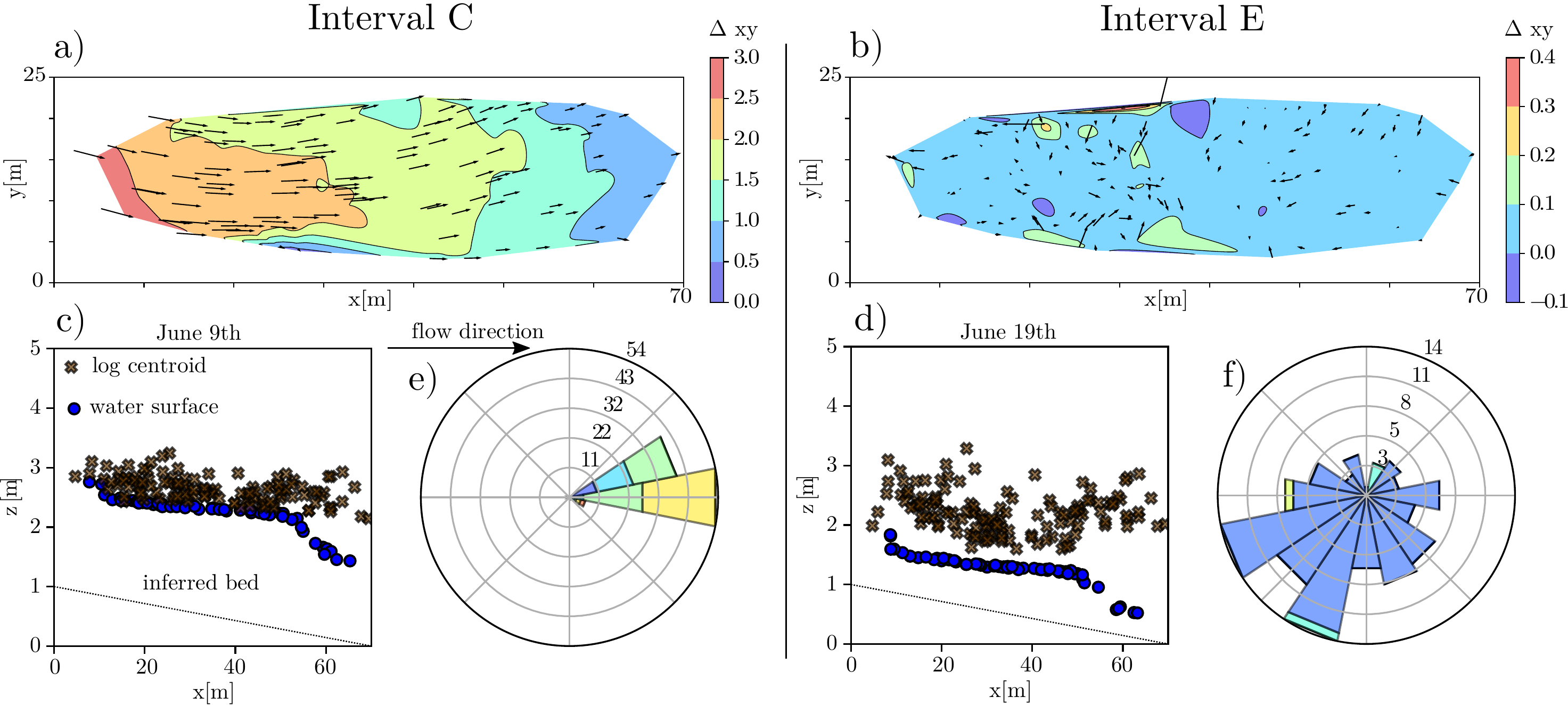}
\caption{\label{fig:arrows_profiles}\textbf{Deformation patterns during intervals C and E.} Interpolated fields of horizontal logjam deformation with arrows showing measured log displacement vectors (a and b) and wind rose diagrams showing the distribution of magnitudes and directions of displacements (e and f).  Longitudinal sections (c and d) show the measured water surface and log midpoints at the beginning of each interval.}
\end{figure}

\subsection{Time-Lapse Photography}
Time-lapse photos provide a more continuous, qualitative record of displacement observations that compliment precise yet discrete measurements from the total station. The photos reveal that the logjam `breathes' - rising vertically during the evening hours and advancing downstream with each diel peak in stage. As stage recedes, the logjam decreases in elevation and downstream  motion decreases but is still present. Multiple perspectives of the logjam demonstrate the closed nature of the system: no logs are lost from the front and no logs are added from upstream. The fabric of the logjam is also preserved, there is little rotation of logs nor locked domains bound by faults or shear bands. 

\pagebreak{}

\section{Discussion}

\subsection{Logjam deformation partitions into three phases}
Logjam deformation is linked to the rising limb, peak, and falling limb of the hydrograph. Small magnitude, downstream displacements occur during the rising limb, amplified by diel snowmelt-driven pulses in stage. These collective, downstream-oriented ensemble movements are greatest at the peak of the hydrograph. As water level falls, the logjam settles in kind; log motions are discrete, low in magnitude and randomly oriented. We infer that these three phases of logjam movement correspond to two modes of deformation: shortening during compression and random displacements during settling. Quantitatively distinct behaviors on both limbs of the hydrograph are demonstrative of non-linearity in the physical response of the logjam to hydrodynamics.

\subsection{Competition between shear, buoyancy and pinning}
Compression can be accommodated by increasing the volume fraction horizontally, and/or by stacking logs vertically. This occurs only during peak flows in the front of the logjam, near the hillslope deposit which inhibits the downstream progression of wood and therefore `pins' the jam  (Figure \ref{hydrograph}). Horizontal compression is ultimately a result of fluid shear, while vertical deformation is a function of stage rise and fall and the buoyant response of logs to water. We infer that these two irreversible processes are critically linked. When buoyant lift is significant enough to release constraining stresses on intertwined or grounded logs, it reduces friction and enables them to respond to shear. As water level lowers, so do the logs until some critical depth where logs re-establish frictional contact with the bed. At this point, logs are no longer buoyant but are supported by the bed, banks and each other, thus inhibiting downstream deformation. Our kinematic measurements result from the competition between shear and buoyancy, and while qualitative, help bridge previously proposed conceptual force balances \citep{Wohl2011,Manners2008} with a fully dynamical model of log interactions and logjam deformation.

Preservation of the `herringbone' geometrical structure of the logjam and the absence of locked domains is in contrast to qualitative observations from historical photos of logjams \citep{Borzsonyi2013} and experiments of horizontal pattern formation in vibrated rods \citep{Narayan2006,Galanis2005,Kudrolli2008}. This is because the particle length is a control parameter which dictates the patterning of driven elongated particles. In our case, the average log width is much larger than the channel width. Vertical structure within the logjam is driven by increased water levels, which create accommodation space beneath the logjam, and the possibility for logs to be stacked on top of each other. This `heaping' behavior is similar that observed in experiments of sheared elongated particles \citep{Fischer2016}. 

Inflections in displacement and water surface profiles occur at the same location in the logjam ($\sim$50m) - proximate to the hillslope debris. We interpret these inflections in xy displacements and water level as being controlled by the pinning debris deposit. Logjam accumulation, growth and persistence clearly depends on channel geometry \citep{Curran2010}, but is also dependent on pinning obstructions in the form of bridge piers \citep{Schmocker2010,Cicco2015,Gschnitzer2015} and by naturally occurring debris jams \citep{Lancaster2006}. Improving our understanding of the interplay between deformation dynamics and the nature of the pinning site could benefit greatly from theoretical investigations \citep{Reichhardt2019} or other clogging studies \citep{Borzsonyi2016}. Our observations lead us to speculate about the mechanisms of logjam break up and propagation downstream: a) destroy obstruction, b) over-ride obstruction, c) circumvent obstruction.

\subsection{Backwater persistence and drag}
At low stage, a portion of logs are submerged and thus are still being subjected to fluid shear. This amount of shear may be non-negligible, as the persistence of the inflection of the water surface profile demonstrates a constant gravitational acceleration of flow and significant drag on the logjam. Large wood effects on channel hydraulics are well-known from field measurements \citep{Abbe1996} and the geometry of our water surface profile is similar to backwater effects studied in flume studies of debris accumulation \citep{Schmocker2013,Schalko2019,Schalko2019b}. Drag coefficients for single wood pieces have been measured in field experiments \citep{Shields2012,Alonso2013}, proposed in numerical models \citep{Hygelund2003,Manga2000} and measured for stable, engineered structures \citep{Bennett2015,Gallisdorfer2014a}, but our measurement of the hydrodynamics are not sufficient to make a full connection with these models. Future work could seek to make explicit force measurements of logs, or to constrain flow velocities, with the intent of making connections with hydrodynamic models of drag. 

\section{Conclusions}
We conducted field investigations to map the spatio-temporal patterns of deformation within in a channel-spanning mountain logjam in Big Creek, Idaho. Our measurements demonstrate that the logjam is not jammed, but is highly sensitive to variations in river flow. Deformation is driven by the competition of shear, buoyancy, and the stability of the hillslope avalanche debris. These observations complicate the task of developing thresholds for logjam break up and downstream propagation, as precisely determined thresholds for movement are memory-dependent and time-varying. Therefore, applications of constitutive models which require the specification of a finite yield stress are problematic. Although our observations are from a single field site, the modes of deformation and sensitivity to hydrograph shape and timing should be applicable to logjams in constrained channels with time varying flow. Further work which explicitly measures forces in the logjam can be more concretely connected to experiments and simulations of granular rheology. The language of pinning, clogging and creep open the space for quantitative treatment of the phenomenology of wood-laden flows and their states of arrest as logjams. This can lead to more informed wood retention structures for restoration, bridge pier design, river corridor management and flood hazard mitigation.

\section{Acknowledgments}
We thank the DeVlieg Foundation for providing funding to support this work. We also thank the US Forest Service and Arnold Aviation, who provided transportation and field support. Further, the caretakers of Taylor Ranch and the residents of Big Creek Village provided much appreciated assistance. We are also deeply grateful to Caitlin Vitale-Sullivan for assistance in the field.

\pagebreak

\section{Works cited}
\bibliography{main}

\pagebreak{}

\section{Supplementary Materials}
\noindent\textbf{Introduction}
Although considered to be static features, we demonstrate with field measurements from Big Creek, Idaho that these features can exhibit a rich kinematic behavior which is tied to the shape and timing of the hydrograph. In the main text, we discuss and explore this coupling. Our supplementary dataset include the hydrologic conditions that we infer to have triggered the avalanche delivery of logs to the stream from the hillslope. Further, we include are a series of high-precision total station measurements which provide six discrete snapshots of logjam deformation. Due to space constraints and for the readability of the paper, we have included the all deformation, water level and log centroid position data within these supplemental materials. We have included a table which summarizes the important features of each survey interval. All data (and Jupyter notebook which analyzed the data and prepared the figures) are being prepared for hosting on our research GitHub repositories.

\noindent\textbf{Time-Lapse Videos}
Time-lapse videos are currently hosted on YouTube at the following links:

\href{https://www.youtube.com/watch?v=F9fSI1LJKIs}{Downstream Perspective} 

\href{https://www.youtube.com/watch?v=iWzza-p4bIY}{Middle Perspective} 

\href{https://www.youtube.com/watch?v=tsHOOuZQ1J8}{Front Perspective}

\begin{center}
\begin{figure}[h]
\noindent\includegraphics[width=1\linewidth]{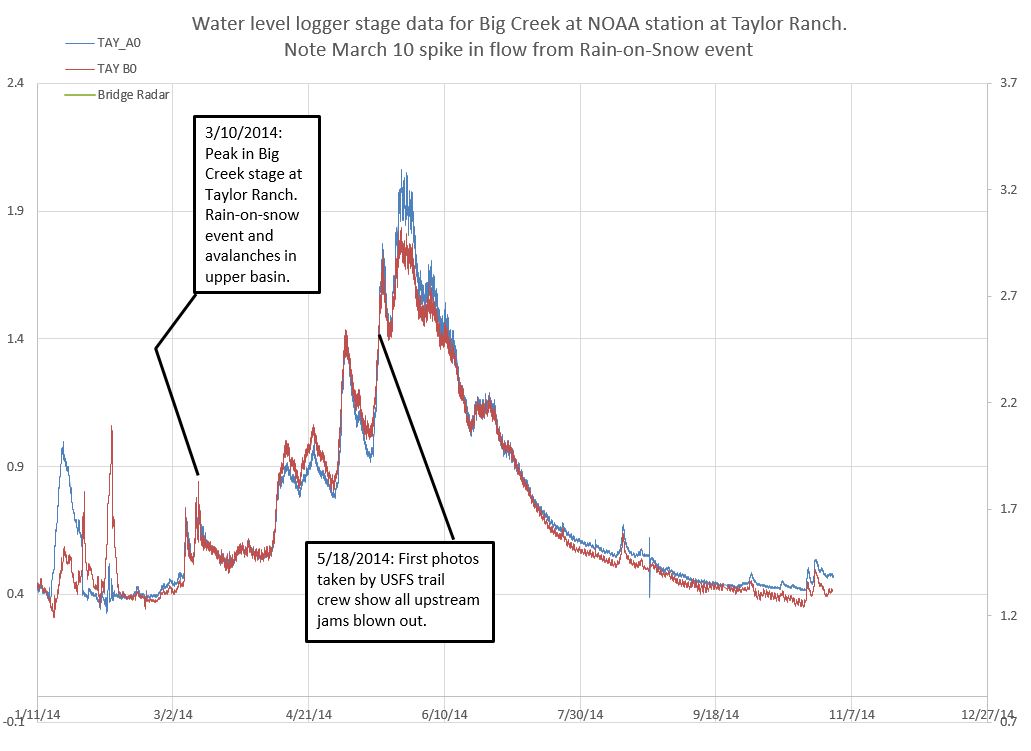}
\caption{\textbf{Hydrograph during spring of 2014 - Taylor Ranch Research Station} The Big Creek watershed is located in the remote Frank Church Wilderness of No Return -- thus direct observations of the logjam formation are not available. Shown here are river stage data during Spring 2014. Boxes indicate the timing of rain-on-snow event which triggered snow avalanches and the first visual evidence of wood-snow debris deposits.}
\label{taylor_ranch_hydrograph}
\end{figure}
\end{center}

\begin{center}
\begin{figure}[h]
\noindent\includegraphics[width=1\linewidth]{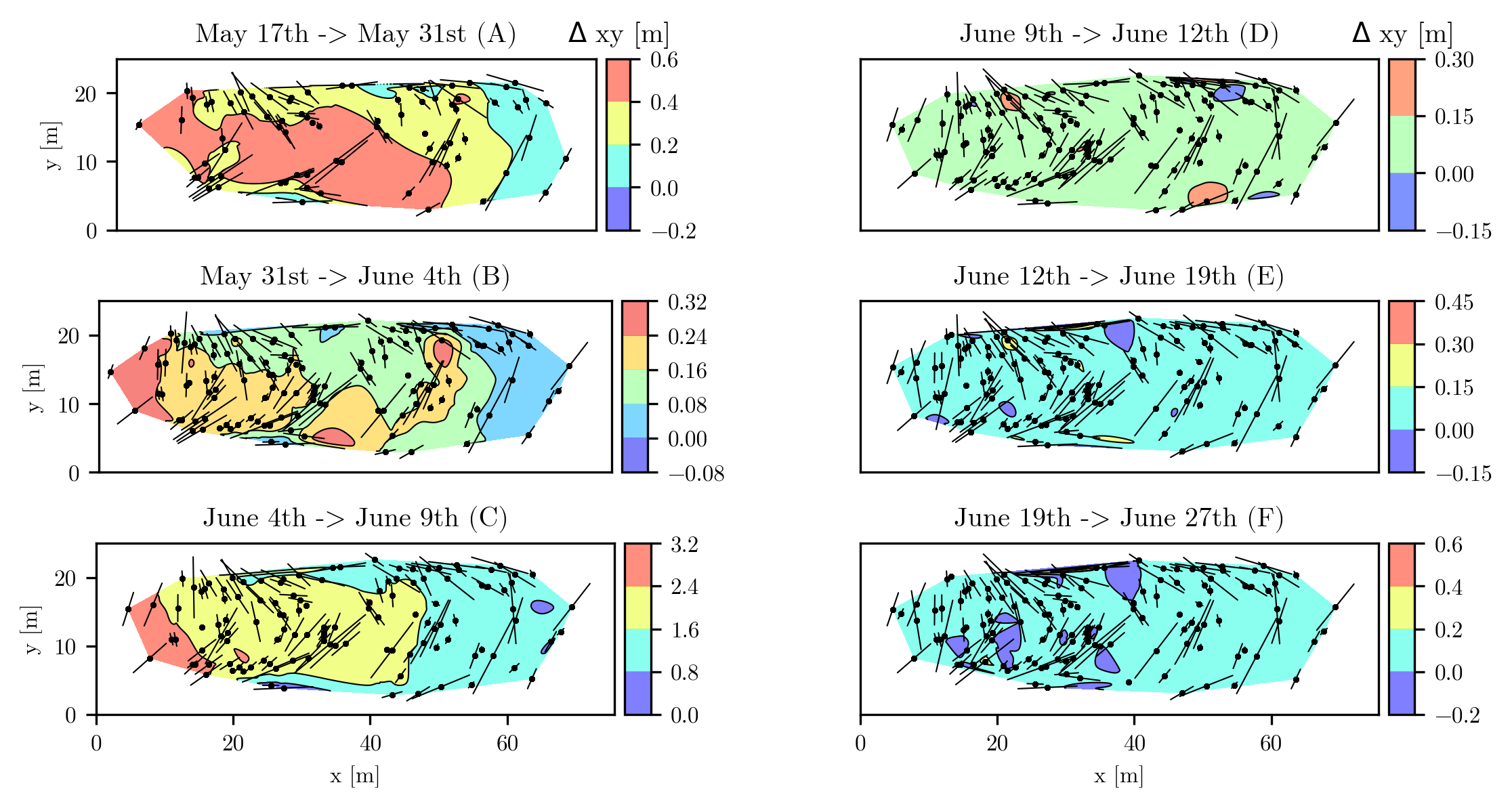}
\caption{\textbf{$\Delta$ xy maps.} Maps of deformation within the Big Creek logjam. Each panel renders log positions (lines), centroids (dots) and a continuous contour map generated via an interpolation of centroid displacements during the survey interval.}
\label{dxy_maps}
\end{figure}
\end{center}

\begin{center}
\begin{figure}[h]
\noindent\includegraphics[width=1\linewidth]{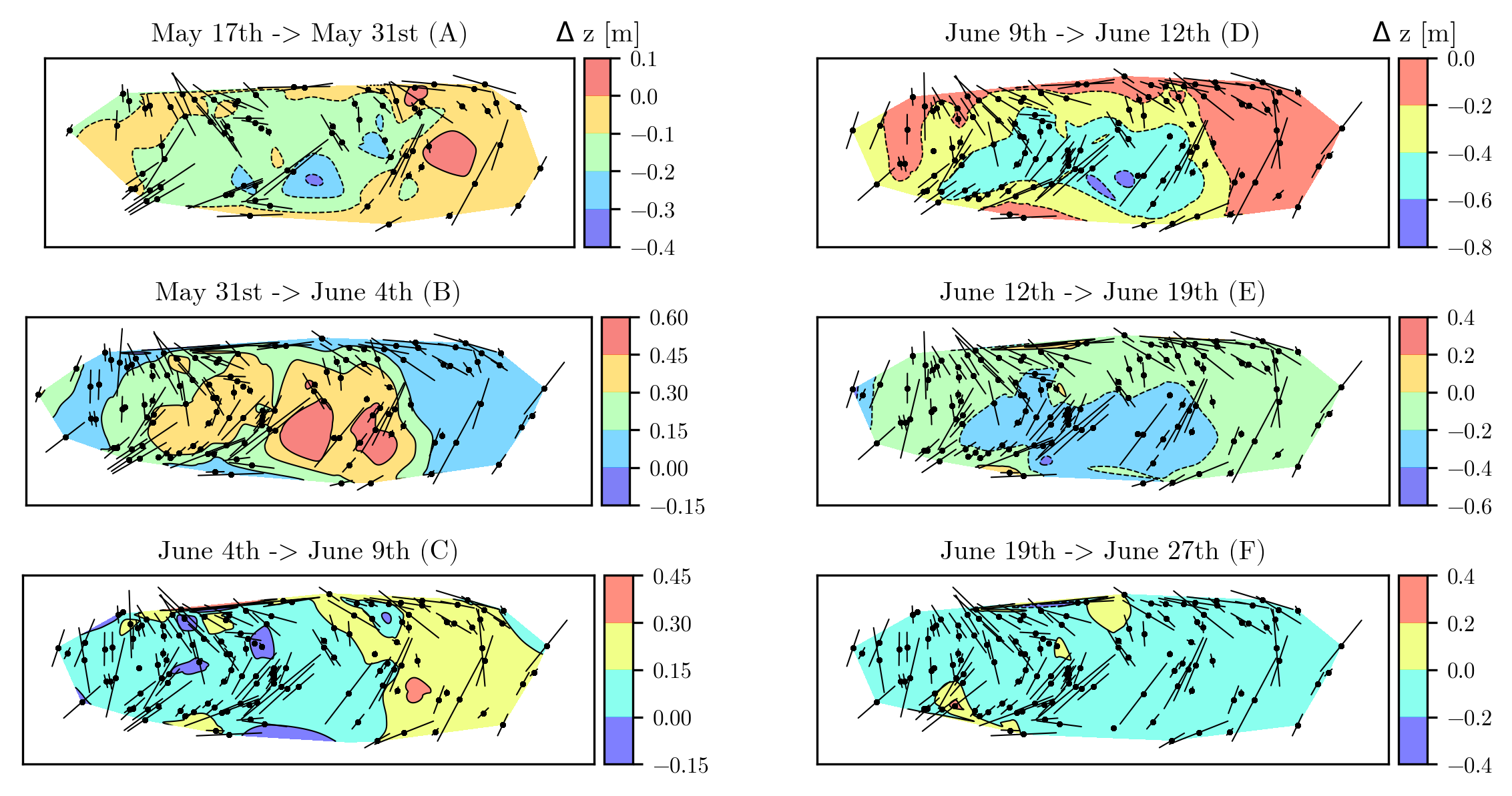}
\caption{\textbf{$\Delta$ z maps.} Maps of deformation within the Big Creek logjam. Each panel renders log positions (lines), centroids (dots) and a continuous contour map generated via an interpolation of centroid displacements during the survey interval.}
\label{dz_maps}
\end{figure}
\end{center}

\begin{center}
\begin{figure}[h]
\noindent\includegraphics[width=1\linewidth]{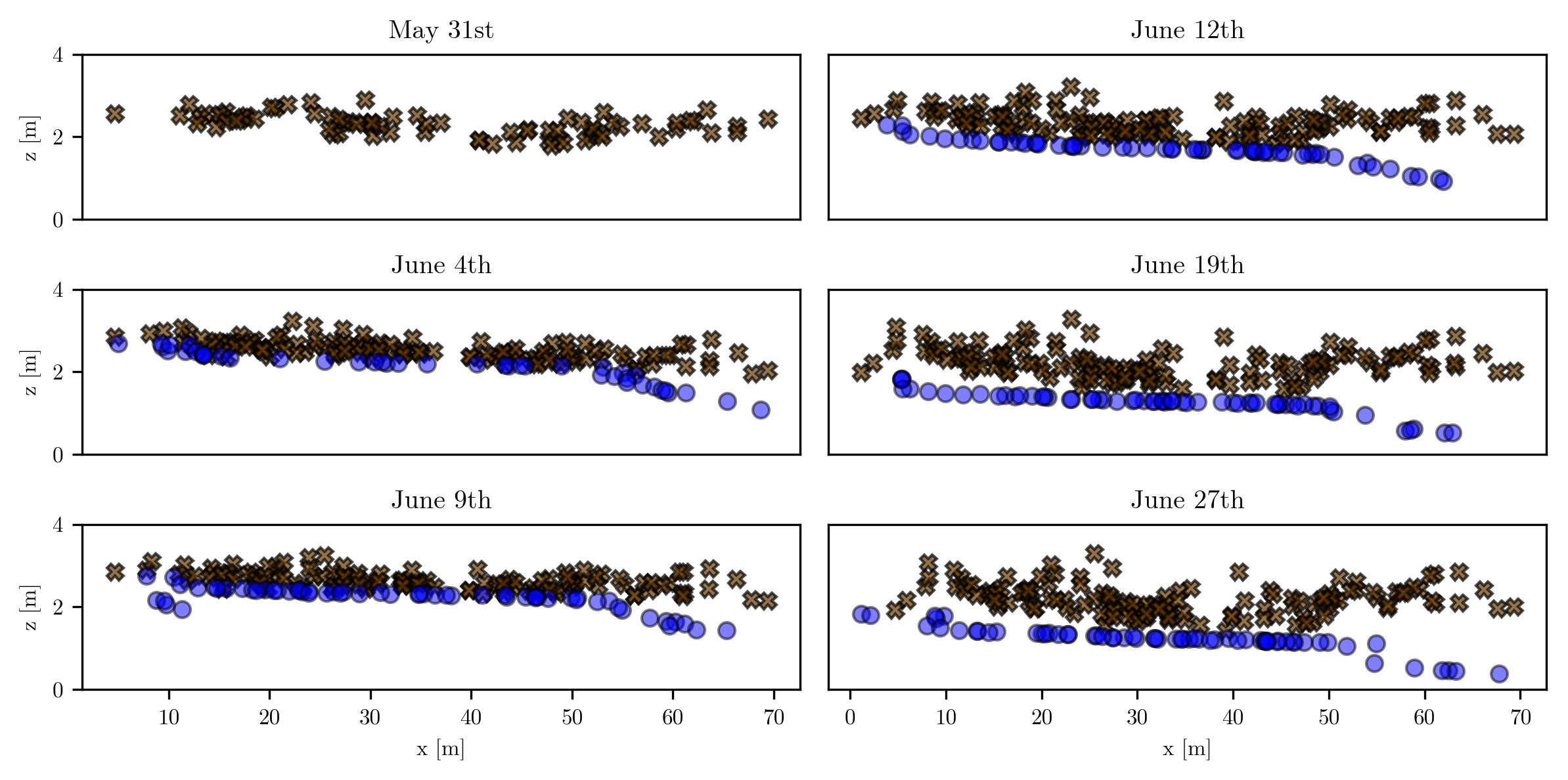}
\caption{\textbf{Water surface profiles.} Log centroid (brown X's) and water surface elevations (blue dots) during a survey, and at the beginning of survey interval (note no water surface data were collected on May 31st). Log centroid elevations cluster tightly during high-flow periods. As water level lowers, log centroids become more scattered. Note the slight inflection in the water surface profiles at 50 m -- this is the approximate location of the intersection of the logjam with the hillslope debris deposit.}
\label{ensemble_water_and_logs}
\end{figure}
\end{center}

\begin{center}
\begin{figure}[h]
\noindent\includegraphics[width=1\linewidth]{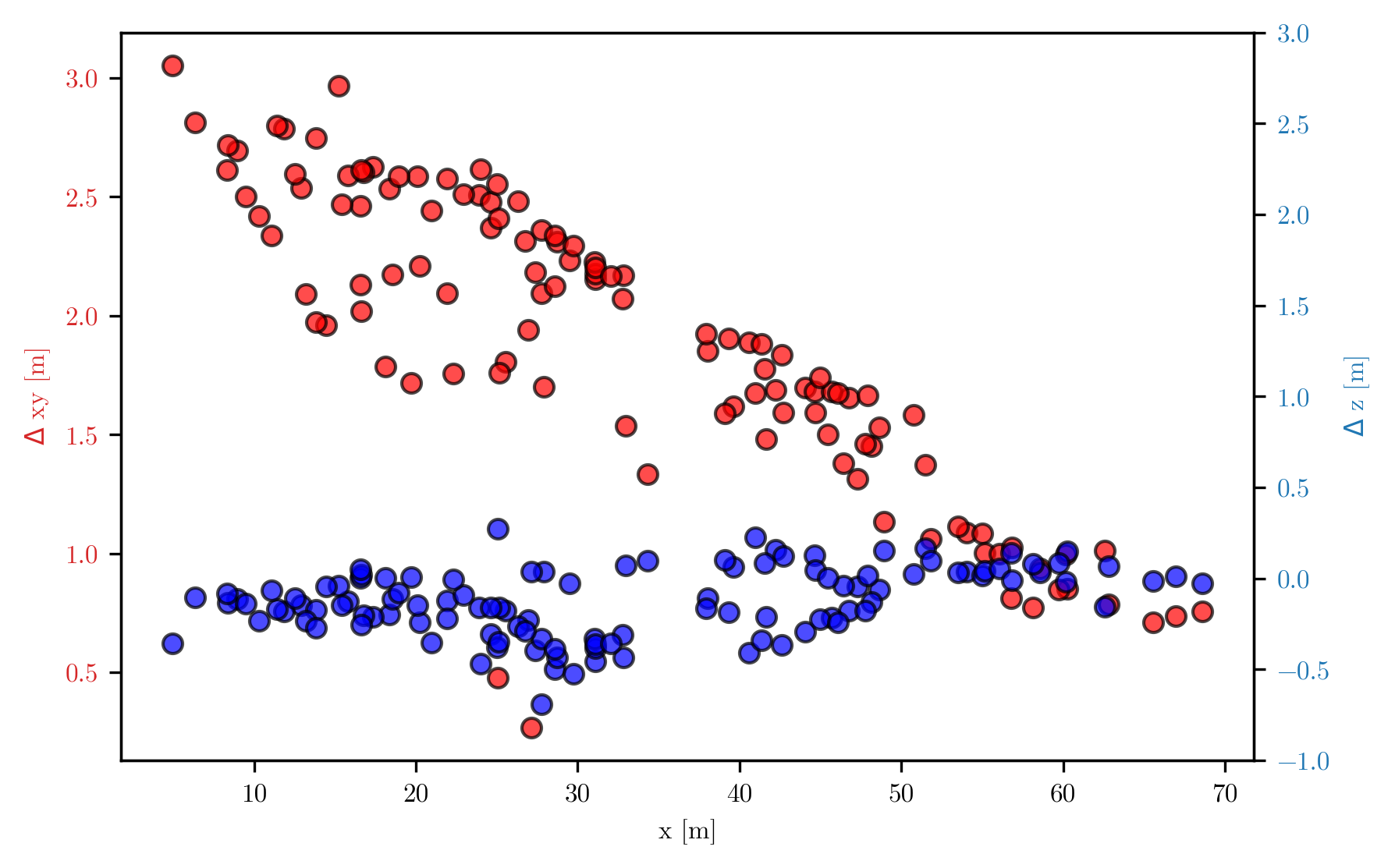}
\caption{\textbf{Net logjam deformation for entire field campaign.} Principle axis (red) shows net horizontal logjam deformation ($\Delta$ xy). Secondary axis (blue) illustrates vertical deformation  ($\Delta$ z). Horizontal displacements are large in magnitude and decay along the distance of the logjam while the vertical deformation is slightly negative.}
\label{total_deformation}
\end{figure}
\end{center}

\begin{sidewaystable}[h]
\caption{Table of survey dates, number of logs and the mean displacements measured.}
\begin{tabular}{|c|c|c|c|c|}
\hline
Survey dates and interval names & \# of surveyed logs & change in stage {[}m{]} & \begin{tabular}[c]{@{}c@{}}mean\\ delta xy {[}m{]}\end{tabular} & \begin{tabular}[c]{@{}c@{}}mean\\ delta z {[}m{]}\end{tabular} \\ \hline
\begin{tabular}[c]{@{}c@{}}May 17th to May 31st\\ (A)\end{tabular} & 73 & -0.119 & 0.325 & -0.099 \\ \hline
\begin{tabular}[c]{@{}c@{}}May 31st to June 4th\\ (B)\end{tabular} & 122 & 0.325 & 0.145 & 0.227 \\ \hline
\begin{tabular}[c]{@{}c@{}}June 4th to June 9th\\ (C)\end{tabular} & 119 & 0.093 & 1.70 & 0.116 \\ \hline
\begin{tabular}[c]{@{}c@{}}June 9th to June 12th\\ (D)\end{tabular} & 130 & -0.532 & 0.067 & -0.290 \\ \hline
\begin{tabular}[c]{@{}c@{}}June 12th to June 19th\\ (E)\end{tabular} & 134 & -0.411 & 0.042 & -0.153 \\ \hline
\begin{tabular}[c]{@{}c@{}}June 19th to June 27th\\ (F)\end{tabular} & 131 & \begin{tabular}[c]{@{}c@{}}n/a\\ (water level fell below installed logger)\end{tabular} & 0.016 & -0.025 \\ \hline
\end{tabular}
\end{sidewaystable}

\end{document}